\documentclass[%
 reprint,
 amsmath,amssymb,
prl,letterpaper,english,
showkeys,showpacs]{revtex4-1}
\usepackage{color}
\usepackage{amsthm}
\usepackage{hyperref}
\definecolor{xblue}{rgb}{0.1,0.6,1}
\hypersetup{
    colorlinks = true,
    allcolors= xblue
    }
    
\usepackage{graphicx}
\usepackage{dcolumn}
\usepackage{bm}

\begin{document}

\preprint{}

\textit{\href{http://dx.doi.org/10.1016/j.scriptamat.2014.11.022}{R. Rahemi, D. Li, Scripta Materialia \textbf{99},41-44 (2015).} }

\title[]{Variation in electron work function with temperature and its effect on the Young's modulus of metals}

\author{Reza Rahemi} 
\author{Dongyang Li }

\affiliation{%
Department of Chemical and Materials Engineering, University of Alberta, Edmonton, Alberta, Canada, T6G 2V4}

\begin{abstract}
Properties of metals are fundamentally determined by their electron behavior, which is largely reflected by the electron work function ($\varphi $). Recent studies have demonstrated that many properties of metallic materials are directly related to $\varphi $, which may provide a simple but fundamental parameter for material design. Since material properties are affected by temperature, in this article a simple model is proposed to correlate the work function with temperature, expressed as $\varphi (T)=\varphi _{0} -\gamma \frac{(k_{B} T)^{2} }{\varphi _{0} } $, where $\gamma $ varies with the crystal structure. This $\varphi $-T relationship helps determine and understand the dependence of metal properties on temperature on a feasible electronic base. As a sample application, the established relationship is applied to determine the dependence of Young's modulus of metals on temperature. The proposed relationship is consistent with experimental observations. 
  
\end{abstract}


\maketitle

\section{Introduction}
Material properties are fundamentally correlated to the electron behavior, which is largely reflected by the electron work function \cite{1,2,3,4,5,6,7,8}. This correlation with work function includes a number of factors, including the Young's modulus, thermal expansion, and heat capacity \cite{9,10,11}. Recent studies \cite{10,11,12} have shown that it may be more feasible to use the work function in material design compared to relevant quantum theories, since the latter are rather difficult to apply in material design, especially for structural materials.

Many properties of materials are strongly affected by temperature. This is probably related to the influence of temperature on the behavior of electrons. The main objective of this work is to establish a relationship between the work function and temperature, so that the dependence of the material properties on temperature can be predicted via the effect of temperature on the work function, which also helps our fundamental understanding of such dependence. With the established $\varphi -T$ relationship, we have predicted the dependence of Young's modulus on temperature as a sample application.

Regarding the effect of temperature on Young's modulus, a model to describe the variation in elastic modulus with temperature was first proposed by Born and Huang in 1954 \cite{13}, in which the Temperature dependence of elastic modulus results from non-harmonic changes in lattice potential energy, originally caused by lattice vibrations \cite{13,14}. They demonstrated that the Young's modulus varied with $T^{4} $. Although consistent with the third law of thermodynamics, experimental results show that this dependence is quite limited as it is only valid when the temperature approaches absolute zero, which is ultimately negligible when considering larger ranges of temperature (e.g, 100$^{\circ } K$ to 1000$^{\circ } K$) that are more meaningful to engineering applications of metallic materials at various temperatures. Wachtman et. al \cite{15} have shown, through experimental data fitting, that Young's modulus may be described as

\begin{equation} \label{GrindEQ__1_} 
E=E_{0} -BTexp(-T_{0} /T) 
\end{equation}

\noindent where $E_{0} $, $B$ and $T_{0} $ are empirical constants. However, such an empirical equation does not provide a clear mechanism for the described relationship. Besides, Anderson \cite{16} has shown that this equation is only valid when the variation of the Poisson ratio with temperature is small.

\begin{figure}
\scalebox{0.28}{\includegraphics*{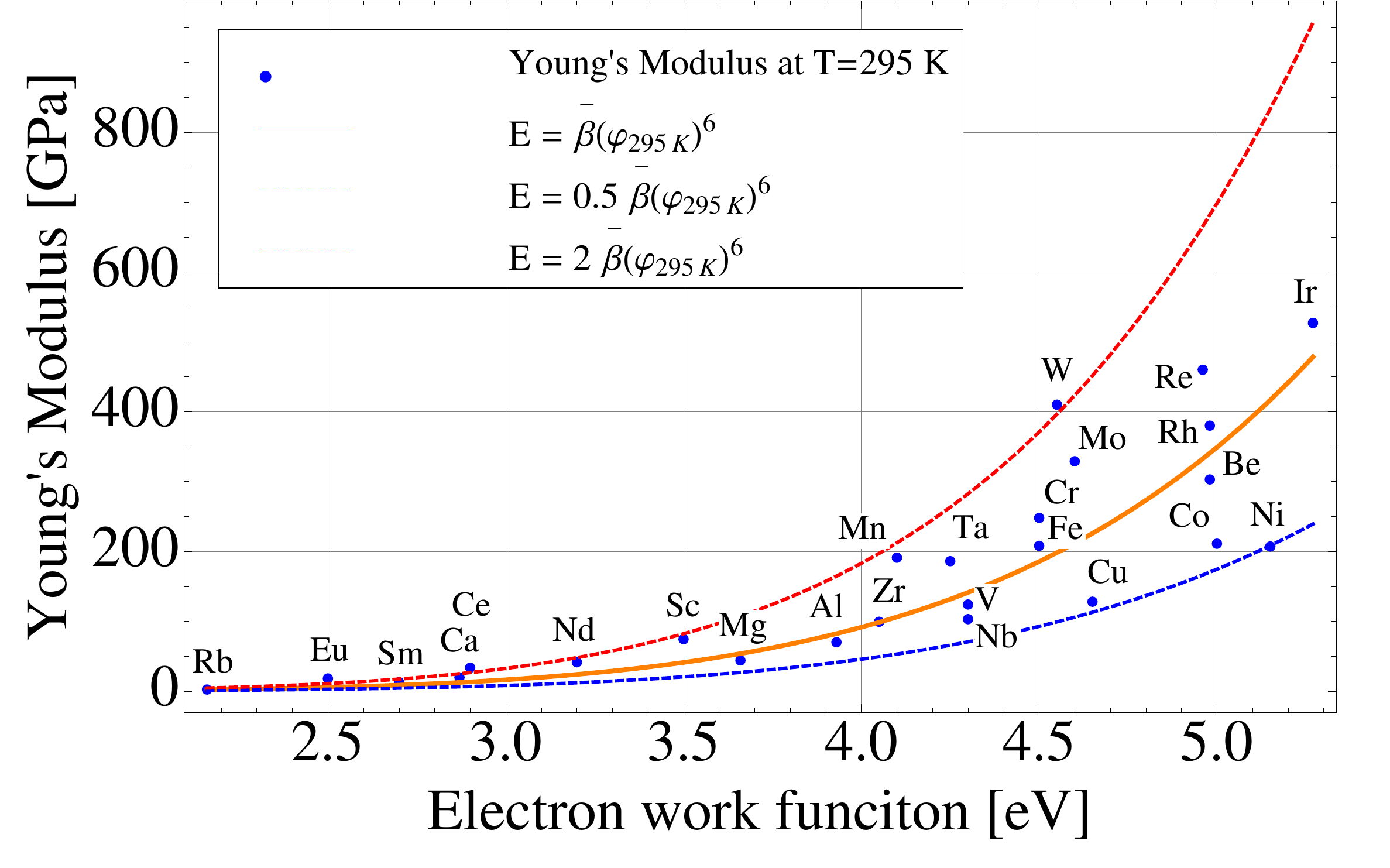}}
\caption{Correlation between electron work function and Young's modulus of metals (adapted from reference \cite{9}). The collected experimental data points are around an average curve of $E=\bar{\beta }\varphi ^{6} $ with $\bar{\beta }=0.02233\frac{GPa}{eV^{6} } $. The upper and lower bounds of the band, within which all the data points are located, correspond to $\beta $ equal to $2\bar{\beta }$ and $0.5\bar{\beta }$, respectively.}

\label{fig1}
\end{figure}

A recent study by Hua and Li \cite{9} has correlated the Young's modulus of metals to their work function, with which the dependence of Young's modulus on temperature could be predicted if the effect of temperature on work function can be established. As demonstrated, the Young's modulus of metals, $E$, has a sextic relation with the work function expressed as

\begin{equation} \label{GrindEQ__2_} 
E=\alpha e^{2} (\frac{16^{6} \times 18\pi ^{10} \hbar \varepsilon _{0}^{9} a^{3} }{e^{16} m^{3} } )^{6} \varphi ^{6}  
\end{equation} 
where $\alpha $ is Madelung's constant, $a$ is the equilibrium lattice constant, $m$ is the electron mass, $e$ is the elementary charge, and $\varepsilon _{0} $ is the vacuum permittivity. A general expression that correlates the Young's modulus with the work function is given as

\begin{equation} \label{GrindEQ__3_} 
E=\beta \varphi ^{6}  
\end{equation} 
where $\bar{\beta }=0.02233\frac{GPa}{eV^{6} }$ is the average value for various crystal structures \cite{9}. This relationship is illustrated in Fig.\ref*{fig1} for various metals. Alternative curves with different values of $\beta $ for different crystal structures are demonstrated in the same figure. The region outlined by these curves represent the variation of the value for $\beta $ from $0.5\bar{\beta }$ to $2\bar{\beta }$. The value of $\beta $ varies with the crystal structure.

Thus, if the dependence of work function on temperature is established, the effect of temperature on Young's modulus can be predicted. The work function-temperature relationship is of significance not only to Young's modulus but also to other properties such as thermal expansion coefficient and heat capacity, etc. In addition to prediction of the dependence of intrinsic properties of materials on temperature, $\varphi (T)$ could also play an important role in acting as an alternative parameter for material design.

The work function can be regarded as a barrier for the electrons to be moved from inside a solid to a point in vacuum immediately outside the solid surface \cite{17}. Increasing temperature should decrease this potential barrier, since electrons would be thermally excited and easier to be moved as the temperature rises.

Dushman \cite{18} has shown that the electron emission from metals is enhanced with an increase in temperature. This is attributed to the energy absorption by electrons inside a metal, with an assumption that the increase in energy of each electron with temperature is equal to $\frac{3}{2} k_{B} T$ ($k_{B} $is the Boltzmann's constant). Thus, the work function may be lowered as 
\begin{equation} \label{GrindEQ__4_} 
\varphi =\varphi _{0} -\frac{3}{2} k_{B} T 
\end{equation}

\noindent where, $\varphi _{0} $ is the work function at $T=0^{\circ } K$ which is expressed as \cite{9}

\begin{equation} \label{GrindEQ__5_} 
\varphi _{0} =\frac{e^{3} m^{1/2} n^{1/6} }{16^{3} \sqrt{3} \pi ^{5/3} \hbar \varepsilon _{0}^{3/2} }  
\end{equation}

\noindent where $n$ is the free electron density \cite{6}. The term $\frac{3}{2} k_{B} T$ is only the average energy of electrons according to the equipartition theorem, which states that the energy is shared between all accessible degrees of freedom, where each degree of freedom contributes $\frac{1}{2} k_{B} T$ to the average internal energy \cite{19}. Each degree of freedom is a parameter that contributes to the state of a system and can be regarded as a classical harmonic oscillator with energy $k_{B} T$. However the equipartition theorem can only be used when the frequency of each oscillation is less than $\frac{k_{B} T}{\hbar } $\cite{19}. According to Bardeen and Pines \cite{20}, the plasma frequency for electrons (in SI units) in a solid is given by 
\begin{equation} \label{GrindEQ__6_} 
\omega _{p} =\sqrt{\frac{ne^{2} }{m\varepsilon _{0} } }  
\end{equation} 
which is much larger than $\frac{k_{B} T}{\hbar } $ at temperatures where materials are applied. Therefore the equipartition theorem is not applicable for the present case.

According to the free electron model, the total energy of free electrons inside a metal is given by: 
\begin{equation} \label{GrindEQ__7_} 
U(T)=\int _{0}^{\infty } \varepsilon f(\varepsilon )D(\varepsilon )d\varepsilon  
\end{equation}

\noindent where $f(\varepsilon )$ is the Fermi-Dirac distribution, given by

\begin{equation} \label{GrindEQ__8_} 
f(\varepsilon )=\frac{1}{exp(\frac{\varepsilon -\varepsilon _{f} }{k_{B} T} )+1}  
\end{equation} 

\begin{equation} \label{GrindEQ__9_} 
D(\varepsilon )=\frac{\partial N}{\partial \varepsilon }  
\end{equation} 
is the density of states and $N$ is given as

\begin{equation} \label{GrindEQ__10_} 
N=V\frac{(2m\varepsilon )^{3/2} }{3\pi ^{2} \hbar ^{3} }  
\end{equation} 
Equation \eqref{GrindEQ__9_} thus becomes: 
\begin{equation} \label{GrindEQ__11_} 
D(\varepsilon )=\frac{\partial N}{\partial \varepsilon } =\frac{3}{2} \frac{N}{\varepsilon }  
\end{equation}

\noindent The average energy for one electron at temperature $T$ is therefore equal to

\begin{equation} \label{GrindEQ__12_} 
\frac{U(T)}{N} =\frac{3}{2} \int _{0}^{\infty } f(\varepsilon )d\varepsilon =\frac{3}{2} \int _{0}^{\infty } \frac{d\varepsilon }{exp(\frac{\varepsilon -\varepsilon _{f} }{k_{B} T} )+1}  
\end{equation} 
Letting $x=\frac{\varepsilon -\varepsilon _{f} }{k_{B} T} $, the integral in \eqref{GrindEQ__12_} can be evaluated as 
\begin{equation} \label{GrindEQ__13_} 
\frac{3}{2} k_{B} T\mathop{\lim }\limits_{x\to \infty } \int _{0}^{x} \frac{dx}{e^{x} +1} =\frac{3 \ ln(2)}{2} k_{B} T\;  
\end{equation}

The average values calculated using these two approaches (i.e the equipartition theorem and the free electron model) are not appropriate choices for determining the change in work function with temperature, since electrons with the highest energy, rather than the average energy, contribute to the change in work function.

\section{Derivation of $\varphi(T)$ using Lennard Jones Potential}

The potential between two atoms, according to the Lennard-Jones potential is commonly expressed as \cite{10}

\begin{equation} \label{GrindEQ__14_} 
V(r)=\varepsilon _{b} [(\frac{r_{e} }{r} )^{12} -2(\frac{r_{e} }{r} )^{6} ] 
\end{equation}

\noindent where $\varepsilon _{b} $ is the maximum value for the depth of the potential well, $r$ is the distance between the two atoms and $r_{e} $ is the equilibrium distance. It has been shown that the electron work function is related to the bond energy \cite{10}. This relationship is given as

\begin{equation} \label{GrindEQ__15_} 
\varphi (r_{e} )=C\varepsilon _{b}^{1/6}  
\end{equation}

\noindent where $\varphi (r_{e} )$ is the work function at equilibrium and $C$ is a constant of proportionality. Since $V(r_{e} )=-\varepsilon _{b} $,

\begin{equation} \label{GrindEQ__16_} 
V(r_{e} )=-\frac{1}{C} \varepsilon _{b}^{5/6} \varphi (r_{e} ) 
\end{equation}

\noindent Because the atomic displacement or vibration, $\Delta r=r-r_{e} $, is within a small range ($r<1.1\; r_{e} $, otherwise the bond would become unstable \cite{21}), equation \eqref{GrindEQ__16_} can be written as

\begin{equation} \label{GrindEQ__17_} 
V(r)=-\frac{1}{C} \varepsilon _{b}^{5/6} \varphi (r) 
\end{equation}

\noindent Combining equations \eqref{GrindEQ__14_} and \eqref{GrindEQ__17_}, the expression for the work function is finalized as:

\begin{equation} \label{GrindEQ__18_} 
\varphi (r)=-\varphi (r_{e} )[(\frac{r_{e} }{r} )^{12} -2(\frac{r_{e} }{r} )^{6} ] 
\end{equation}

\noindent Letting $x=\frac{<r-r_{e} >}{r_{e} } $, $\varphi (r)$ can be written as

\begin{equation} \label{GrindEQ__19_} 
\varphi (r)=-\varphi _{0} [(1+x)^{-12} -2(1+x)^{-6} ]\approx \varphi _{0} -36x^{2} \varphi _{0}  
\end{equation}

\noindent Since the distance between the two atoms (the bond length) is affected by temperature, an expression for a temperature-dependent work function can be derived using this relationship. Since the change in potential energy is similar to that in work function (see Eqs. (\ref{GrindEQ__14_}) and (\ref{GrindEQ__18_})), the temperature dependent $x$ varies as \cite{10} $x=\frac{1}{r_{e} } \frac{3g}{4f^{2} } k_{B} T$, where we may have $g=252\frac{\varphi _{0} }{r_{e}^{3} } $ and $f=36\frac{\varphi _{0} }{r_{e}^{2} } $. Thus, $x$ becomes

\begin{equation} \label{GrindEQ__20_} 
x=(\frac{7}{48} \frac{k_{B} T}{\varphi _{0} } ) 
\end{equation} 
Combining equations \eqref{GrindEQ__19_} and \eqref{GrindEQ__20_}, we have 

\noindent 
\begin{equation} \label{GrindEQ__21_} 
\varphi (T)=\varphi _{0} -36(\frac{7}{48} \frac{k_{B} T}{\varphi _{0} } )^{2} \varphi _{0}  
\end{equation}

\noindent This expression can be generalized for solids by considering the potential due to the interaction of other adjoining atoms. The second term in equation (\ref{GrindEQ__19_}) after the negative sign can be generally described as:

\begin{multline} \label{GrindEQ__22_} 
36\varphi _{0}\sum _{i=1}^N  w_i(\frac{<r_{i} -r_{e_{1} } >}{r_{e_{i} } } )^{2} =\xi \varphi _{0} (\frac{<r_{1} -r_{e_1} >}{r_{e_{1} } } )^{2}  \\
=\gamma \varphi_0 (\frac{k_{B} T}{\varphi _{0} } )^{2}  
\end{multline}

\noindent where  $<r_{i} -r_{e_{1} } >$ is average deviation or displacement of atom $i$ away from the equilibrium position (chosen as origin). $w_i$ is an energy contribution factor that counts the influence of the $i$th neighbor atom on the square of relative oscillation amplitude, $x^2$. This contribution is distance-dependent. e.g., the contributions from the nearest neighbour, $2^{nd}$ and $3^{rd}$ neighbour atoms are    $w_{r_{e_1}}=1$, $w_{r_{e_2}}=3.1 \times 10^{-2}$ and $w_{r_{e_3}}=2.7 \times 10^{-3}$, respectively (determined based on the Lennard-Jones potentials for $r_e$, $2r_e$, $3r_e$, $\dots$ ). $\varphi (T)$ then becomes

\begin{equation} \label{GrindEQ__23_} 
\varphi (T)=\varphi _{0} -\gamma \frac{(k_{B} T)^{2} }{\varphi _{0} }  
\end{equation}

\noindent Combining equations \eqref{GrindEQ__3_} and \eqref{GrindEQ__23_}, the Young's modulus becomes

\begin{equation} \label{GrindEQ__24_} 
E=\beta [\varphi _{0} -\gamma \frac{(k_{B} T)^{2} }{\varphi _{0} } ]^{6}  
\end{equation}

\noindent Since the values for work function are available at the room temperature (i.e 295 K), the relationship for the Young's modulus can be adjusted as

\begin{equation} \label{GrindEQ__25_} 
E=\beta [\varphi _{295} -\gamma \frac{(k_{B} \tau )^{2} }{\varphi _{295} } ]^{6}  
\end{equation} 
where $\tau =T-295K$.

$\gamma $ in eq. \eqref{GrindEQ__22_} can be written as $\gamma =\xi (\frac{7}{48} )^{2} $ and $\xi $ is:
\begin{equation} \label{GrindEQ__26_} 
\xi =\frac{36}{\chi _{1}^{2} } \sum _{i=1}^N (\frac{<r_{i} -r_{e_{1} } >}{r_{e_{i} } } )^{2}  
\end{equation} 
where $\chi _{1} =\frac{<r_{1} -r_{e_{1} } >}{r_{e_{1} } } =\alpha _{L} T$ and $\alpha _{L} $ is the linear thermal expansion coefficient of the metal. Considering the periodicity of the lattice, we have:

\begin{equation} \label{GrindEQ__27_} 
r_{e_{i} } =r_{e_{1} } +(i-1)a  
\end{equation}

\noindent where $a$ is the lattice parameter and $r_{e_{1} } $ is the nearest neighbour distance, which equals $\frac{a}{\sqrt{2}} $ for face-centred cubic (fcc) crystals and $\frac{\sqrt{3} a }{2} $ for body-centred cubic (bcc) crystals, respectively. Thus, $\xi$ for fcc structures becomes:

\begin{equation} \label{GrindEQ__28_} 
\xi _{FCC} =\frac{36}{\chi _{1}^{2} }\sum _{i=1}^{N} w_i(\frac{\chi _{1} +\sqrt{2} (i-1)}{1+\sqrt{2} (i-1)} )^{2}  
\end{equation}

\noindent Similarly, for bcc structures, we have

\begin{equation} \label{GrindEQ__29_} 
\xi _{BCC} =\frac{36}{\chi _{1}^{2} } \sum _{i=1}^{N} w_i(\frac{\chi _{1} +\frac{2}{\sqrt{3} } (i-1)}{1+\frac{2}{\sqrt{3} } (i-1)} )^{2}  
\end{equation}

\begin{table}[Ht]
\begin{centering}

\begin{ruledtabular}

\begin{tabular}{ccc}

Metal&
$\gamma$&
$\beta \ [\bar{\beta}]$

\tabularnewline
\hline 
\tabularnewline
Al&
583&
0.5
\tabularnewline
Fe&
438&
1.1
\tabularnewline
Ag&
478&
0.6
\tabularnewline
Ni&
318&
0.5
\tabularnewline
Cu&
307&
0.6
\end{tabular}
\end{ruledtabular}
\caption{Calculated values for $\gamma $ and estimated values of $\beta $, from figure 1.}
\par\end{centering}
\label{tab_delT}

\end{table}

\begin{figure}[Ht]
\begin{minipage}[b]{0.45\linewidth}
\centering
\includegraphics[width=\textwidth]{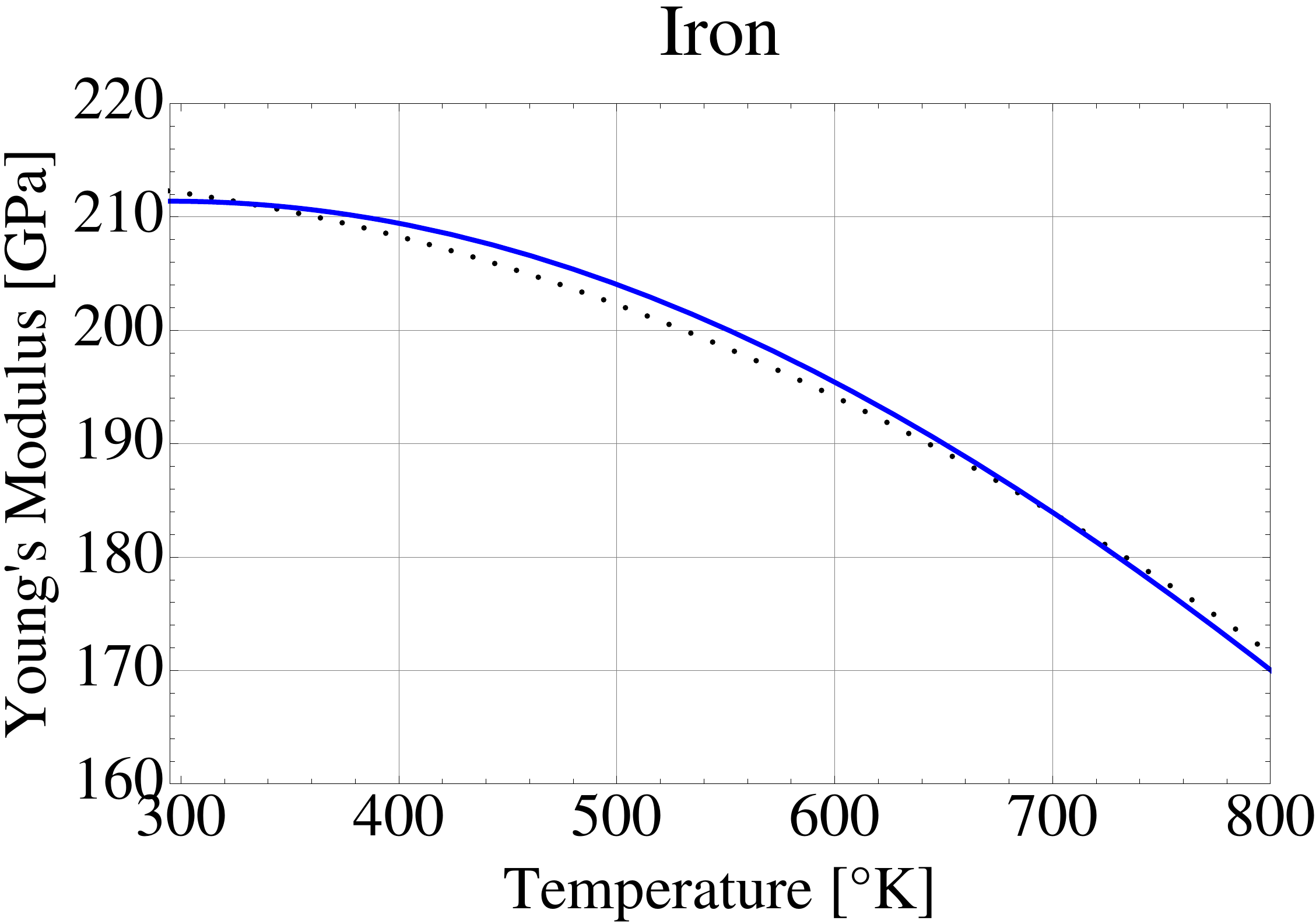}
\end{minipage}
\begin{minipage}[b]{0.45\linewidth}
\centering
\includegraphics[width=\textwidth]{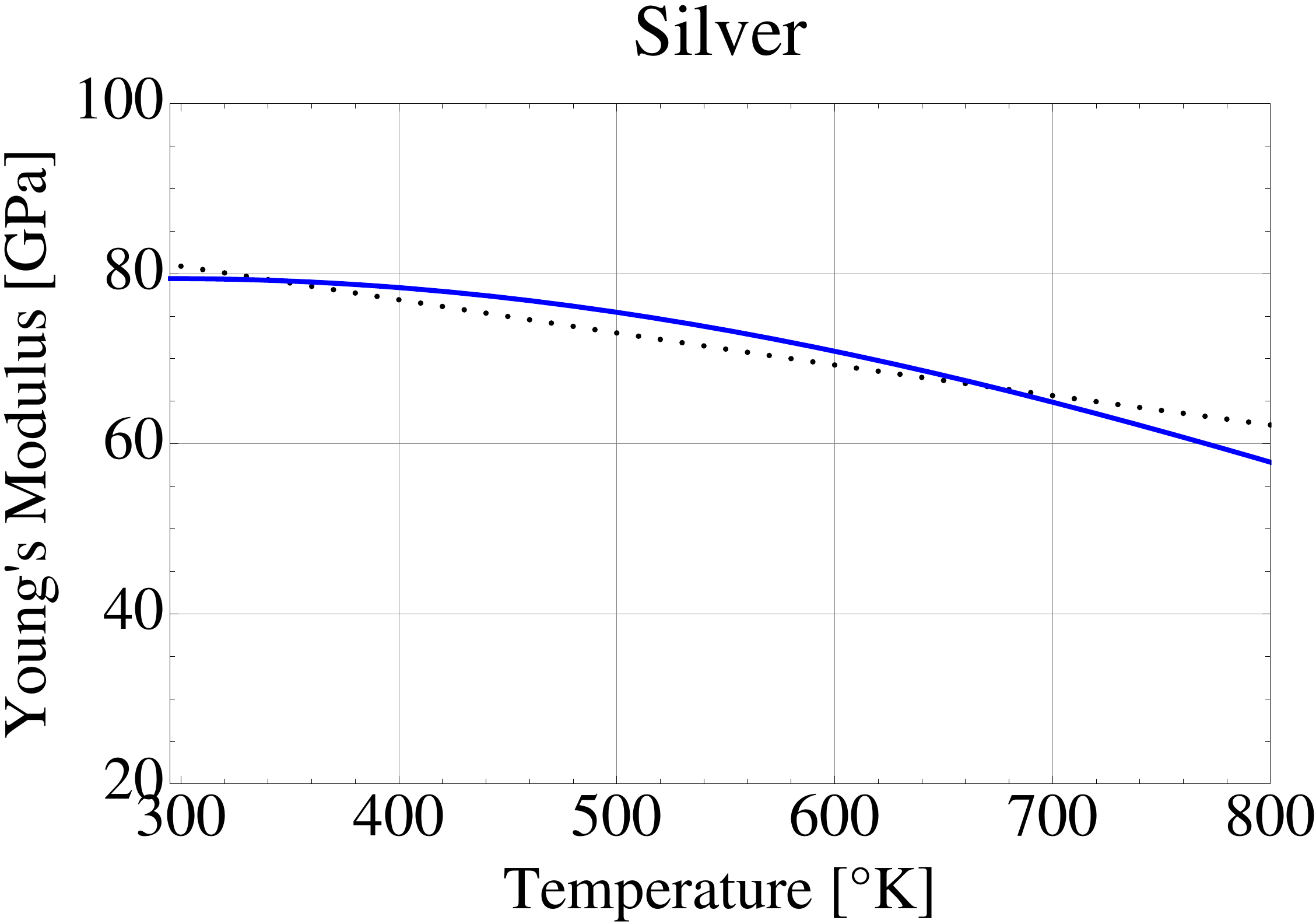}
\end{minipage}
\begin{minipage}[b]{0.45\linewidth}
\centering
\includegraphics[width=\textwidth]{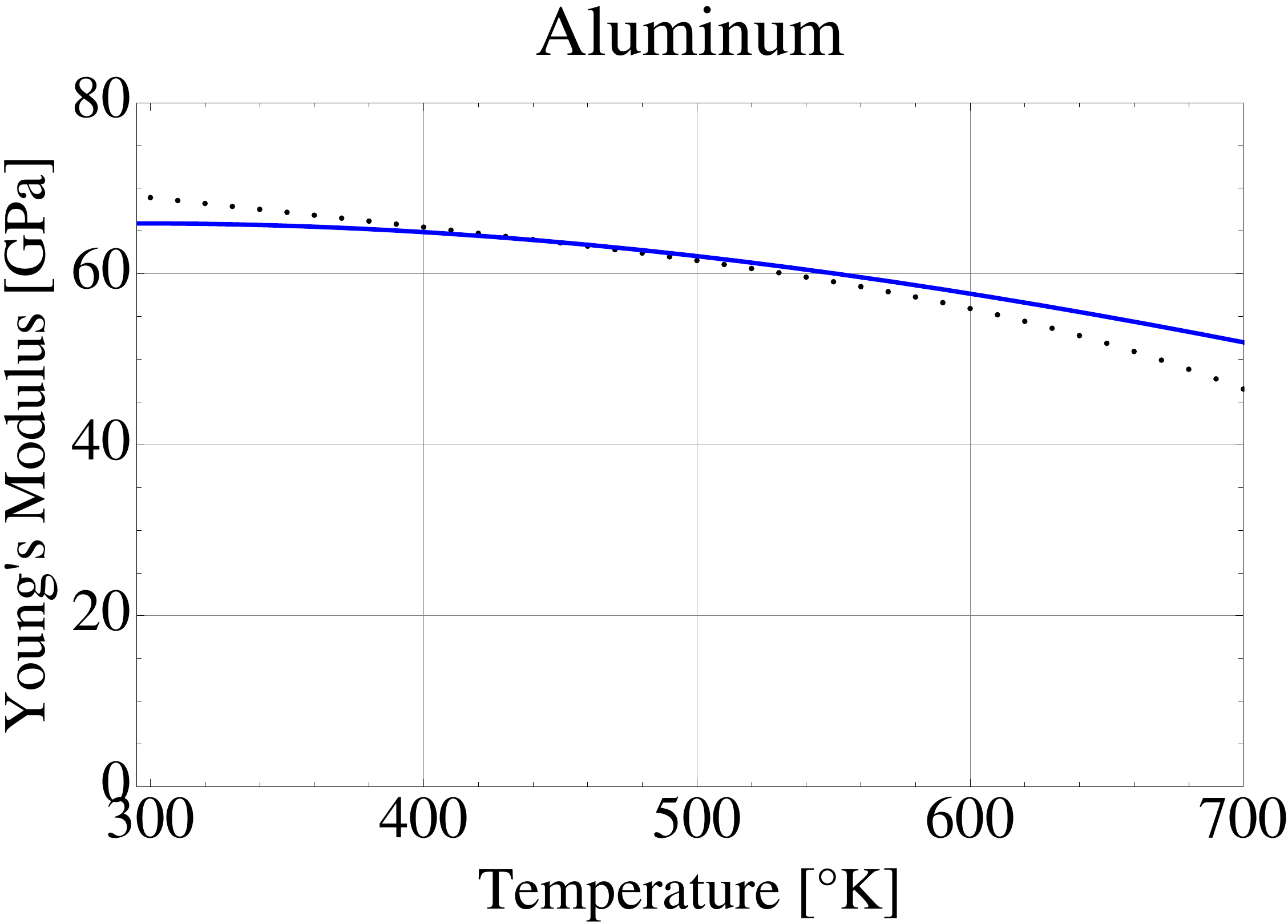}
\end{minipage}
\newpage
\begin{minipage}[b]{0.45\linewidth}
\centering
\includegraphics[width=\textwidth]{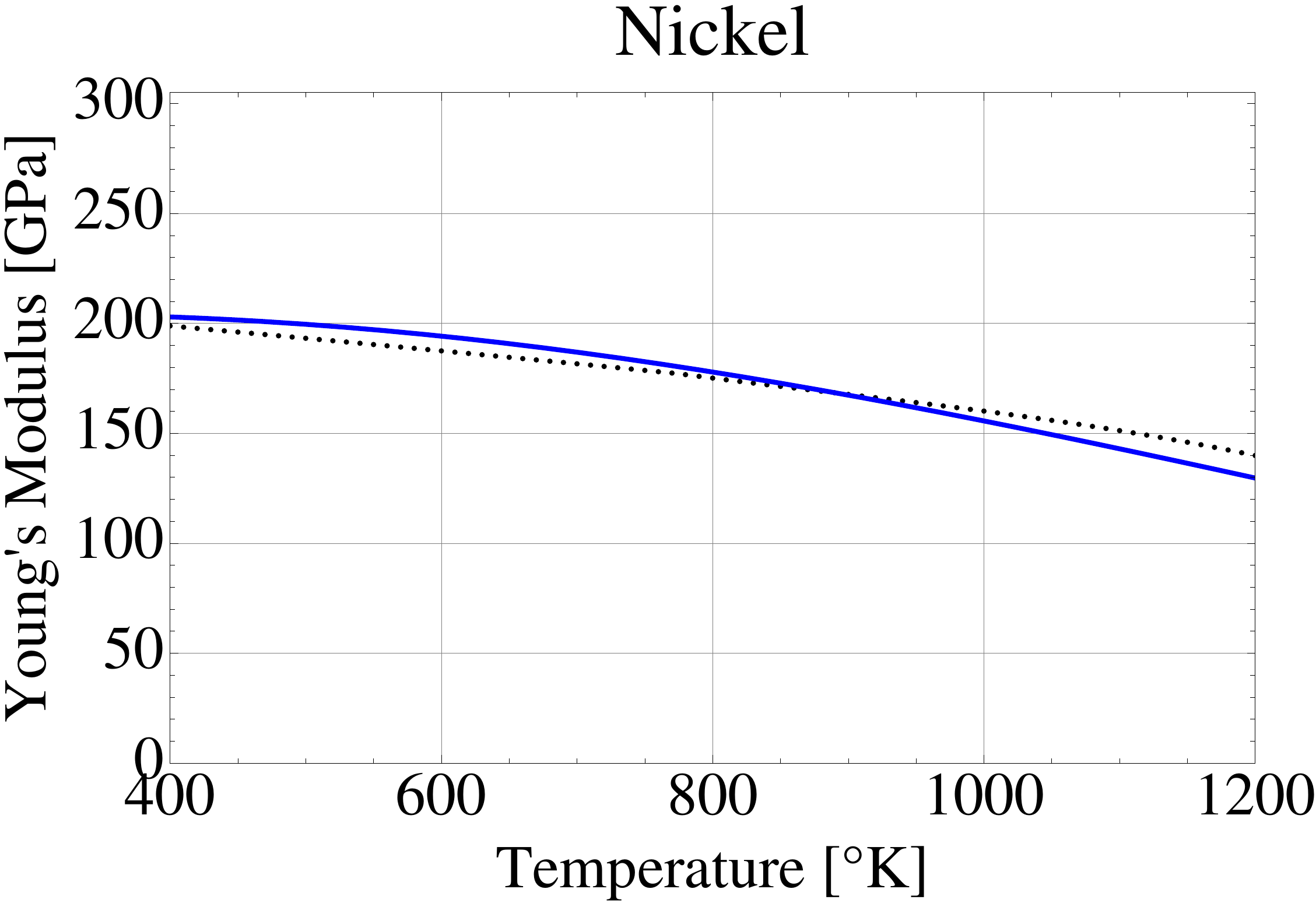}
\end{minipage}
\begin{minipage}[b]{0.45\linewidth}
\centering
\includegraphics[width=\textwidth]{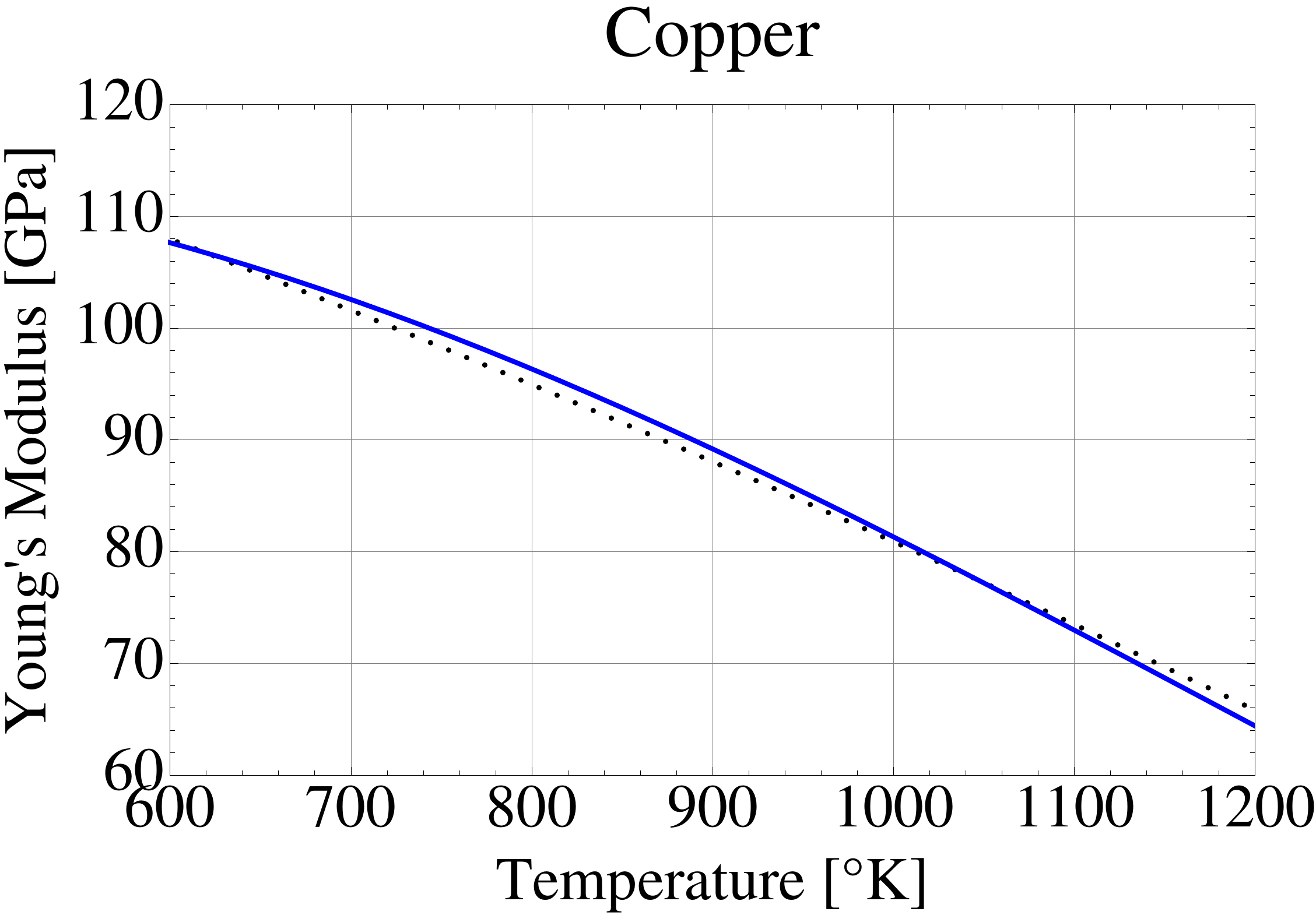}
\end{minipage}

\caption{Variation in Young's modulus of metals with temperature. The solid line represents the prediction and the dashed lines are the experimental values.}\label{multiavp}
\end{figure}

\noindent Considering the interaction between two neighbouring atoms and calculating $\chi _{1} $ at the melting point (the distance between two neighbouring atoms in a metal is maximum at the melting point), the values for $\gamma $ for different metals are calculated and presented in Table 1. With the $\gamma $ values and those for $\beta $, the variation of the Young's modulus with temperature is plotted and presented in Fig. 2. Experimental data are cited from ref. \cite{22,23,24,25,26,27,28,29,30,31,32,33,34,35,36}.

\section{Summary and discussion}
In summary, we propose a relationship between the work function of metals and temperature, $\varphi(T)=\varphi_{0}-\gamma\frac{(k_BT)^2}{\varphi_{0}}$, where the coefficient $\gamma$ is dependent on the crystal structure. Based on this relationship, the temperature dependence of Young's modulus is established. Using iron, silver, aluminum, nickel and copper as examples, variations in their Young's moduli with temperature were predicted, which are supported by reported experimental observations. The $\varphi$-$T$ relationship is of significance not only to Young's modulus but also helps predict the dependence of other intrinsic properties of metals on temperature on a feasible electronic base. The proposed relationship is general, since there is no specific assumption, which could limit the applicability of the relationship, required in the derivation.

\acknowledgments
The authors are grateful for financial support from the Natural Sciences and Engineering Research Council of Canada (NSERC) and are thankful to Dr. Guomin Hua for fruitful discussions.

\end{document}